\documentclass[%
aps,
 amsmath,amssymb,
 superscriptaddress,
reprint,%
longbibliography
]{revtex4-1}

\usepackage{graphicx}% Include figure files
\usepackage{dcolumn}% Align table columns on decimal point
\usepackage{bm}% bold math
\usepackage{ulem}
\usepackage[utf8]{inputenc}
\usepackage[T1]{fontenc}
\usepackage{mathptmx}
\usepackage{textcomp} % for \textsuperscript, \textmu and many more
\usepackage{nicefrac}
\usepackage{xcolor}
\usepackage{gensymb}
\usepackage{multibib}
\usepackage{lineno}
\begin{document}

%\preprint{AIP/123-QED}

\title[Formation of moiré interlayer excitons in space and time]{Formation of moiré interlayer excitons in space and time}

% Force line breaks with \\

\author{David Schmitt} %
\address{I. Physikalisches Institut, Georg-August-Universit\"at G\"ottingen, Friedrich-Hund-Platz 1, 37077 G\"ottingen, Germany}

\author{Jan Philipp Bange} %
\address{I. Physikalisches Institut, Georg-August-Universit\"at G\"ottingen, Friedrich-Hund-Platz 1, 37077 G\"ottingen, Germany}

\author{Wiebke Bennecke} %
\address{I. Physikalisches Institut, Georg-August-Universit\"at G\"ottingen, Friedrich-Hund-Platz 1, 37077 G\"ottingen, Germany}

\author{AbdulAziz AlMutairi} 
\address{Department of Engineering, University of Cambridge, Cambridge CB3 0FA, U.K.}

\author{Kenji Watanabe} %
\address{Research Center for Functional Materials, National Institute for Materials Science, 1-1 Namiki, Tsukuba 305-0044, Japan}

\author{Takashi Taniguchi} %
\address{International Center for Materials Nanoarchitectonics, National Institute for Materials Science, 1-1 Namiki, Tsukuba 305-0044, Japan}

\author{Daniel Steil} %
\address{I. Physikalisches Institut, Georg-August-Universit\"at G\"ottingen, Friedrich-Hund-Platz 1, 37077 G\"ottingen, Germany}

\author{D. Russell Luke} %
\address{Institute for Numerical and Applied Mathematics, Georg-August-Universit\"at G\"ottingen, Lotzestrasse 16-18, 37083 G\"ottingen, Germany}

\author{R. Thomas Weitz} %
\address{I. Physikalisches Institut, Georg-August-Universit\"at G\"ottingen, Friedrich-Hund-Platz 1, 37077 G\"ottingen, Germany}

\author{Sabine Steil} 
\address{I. Physikalisches Institut, Georg-August-Universit\"at G\"ottingen, Friedrich-Hund-Platz 1, 37077 G\"ottingen, Germany}

\author{G.~S.~Matthijs~Jansen} %
\address{I. Physikalisches Institut, Georg-August-Universit\"at G\"ottingen, Friedrich-Hund-Platz 1, 37077 G\"ottingen, Germany}

\author{Stephan Hofmann} 
\address{Department of Engineering, University of Cambridge, Cambridge CB3 0FA, U.K.}

\author{Marcel Reutzel} \email{marcel.reutzel@phys.uni-goettingen.de}%
\address{I. Physikalisches Institut, Georg-August-Universit\"at G\"ottingen, Friedrich-Hund-Platz 1, 37077 G\"ottingen, Germany}

\author{Stefan Mathias} \email{smathias@uni-goettingen.de}%
\address{I. Physikalisches Institut, Georg-August-Universit\"at G\"ottingen, Friedrich-Hund-Platz 1, 37077 G\"ottingen, Germany}

\begin{abstract}

%%%%%%%%%%%%%%%%%%%%%%%%%%%%%%%%%%%%%%%%%%%%%%
%%%%%%%%%%%%% referenced summary paragraph nature style
%%%%%%%%%%%%%%%%%%%%%%%%%%%%%%%%%%%%%%%%%%%%%%

Moiré superlattices in atomically thin van-der-Waals heterostructures hold great promise for an extended control of electronic and valleytronic lifetimes~\cite{Hong14natnano, Rivera15natcom, Rivera16sci, Kim17sciadv, Wang17prb, Merkl19natmat, Ovesen19comphys,Forg21natcom}, the confinement of excitons in artificial moiré lattices~\cite{Hongyi17sciadv, Wu18prb, Seyler19nat, Alexeev19nat, Tran19nat}, and the formation of novel exotic quantum phases~\cite{Su08natphys, Wu18prl, Cao18nat, Cao18nat2, Wang20natmat, Regan20nat}. Such moiré-induced emergent phenomena are particularly strong for interlayer excitons, where the hole and the electron are localized in different layers of the heterostructure~\cite{Rivera18natnano, Jin18natnano, Jiang21light}. In order to exploit the full potential of correlated moiré and exciton physics, a thorough understanding of the ultrafast interlayer exciton formation process and the real-space wavefunction confinement in the moiré potential is indispensable. However, direct experimental access to these parameters is limited since most excitonic quasiparticles are optically dark. Here we show that femtosecond photoemission momentum microscopy provides quantitative access to these key properties of the moiré interlayer excitons. We find that interlayer excitons are dominantly formed  on the sub-50~fs timescale via interlayer tunneling at the K valleys of the Brillouin zones. In addition, we directly measure energy-momentum fingerprints of the moiré interlayer excitons by mapping their spectral signatures within the mini Brillouin zone that is built up by the twisted heterostructure. From these momentum-fingerprints, we gain quantitative access to the modulation of the exciton wavefunction within the moiré potential in real-space. Our work provides the first direct access to the interlayer moiré exciton formation dynamics in space and time and reveals new opportunities to study correlated moiré and exciton physics for the future realization of exotic quantum phases of matter.

\end{abstract}

\maketitle

%%%%%%%%%%%%%%%%%%%%%%%%%%%%%%%%%%%%%%%%%%%%%%
%%%%%%%%%%%%% Introduction
%%%%%%%%%%%%%%%%%%%%%%%%%%%%%%%%%%%%%%%%%%%%%%

The advent of two-dimensional van-der-Waals materials~\cite{Novoselov12nat, Xu13cr} has led to remarkable new strategies to manipulate correlated material properties on the nanometer length- and the femtosecond time-scale. In transition metal dichalcogenides (TMDs), the exceptionally strong light-matter coupling and the weak Coulomb screening of photoexcited electron-hole pairs allows novel spin, valley, and excitonic properties of matter~\cite{Wang18rmp}. Even more intriguing material properties can be realized in TMDs by stacking two or more monolayers into heterostructures~\cite{Geim11nat, Rivera18natnano, Jin18natnano, Jiang21light}: In type II band aligned TMD stacks, novel excitonic states can be created where the electron and the hole contribution to the exciton are separated between the van-der-Waals-coupled TMDs (Fig. 1d). A key question that to date remains unanswered is how these interlayer excitons (ILX) are formed, i.e., if the charge separation results from interlayer tunneling and subsequent phonon cooling at the K valleys~\cite{Merkl19natmat, Ovesen19comphys}, or if exciton-phonon scattering via the $\Sigma$- and $\Gamma$-valleys leads to charge separation~\cite{Wang17prb, Kunstmann18natphys, Wallauer20prb}. Furthermore, the lattice mismatch and the twist angle between the TMDs induce a moiré superlattice that modulates the potential energy landscape by more than 100~meV~\cite{Tran19nat}, which makes it necessary to understand how precisely the interaction of the exciton and the moiré potential determine the material properties. Most intriguingly, it has been shown that ILXs can be confined within the moiré potential minima, which is the first step towards the realization of artificial quantum dot arrays~\cite{Hongyi17sciadv, Wu18prb, Seyler19nat, Alexeev19nat, Tran19nat} that can host strongly correlated phases~\cite{Su08natphys, Wu18prl, Cao18nat, Cao18nat2, Wang20natmat, Regan20nat}. In this context, a significant open challenge is the experimental quantification of the excitonic wavefunction localization and modulation within the moiré potential.

\begin{figure*}[hbt!]
    \centering
    \includegraphics[width=1\linewidth]{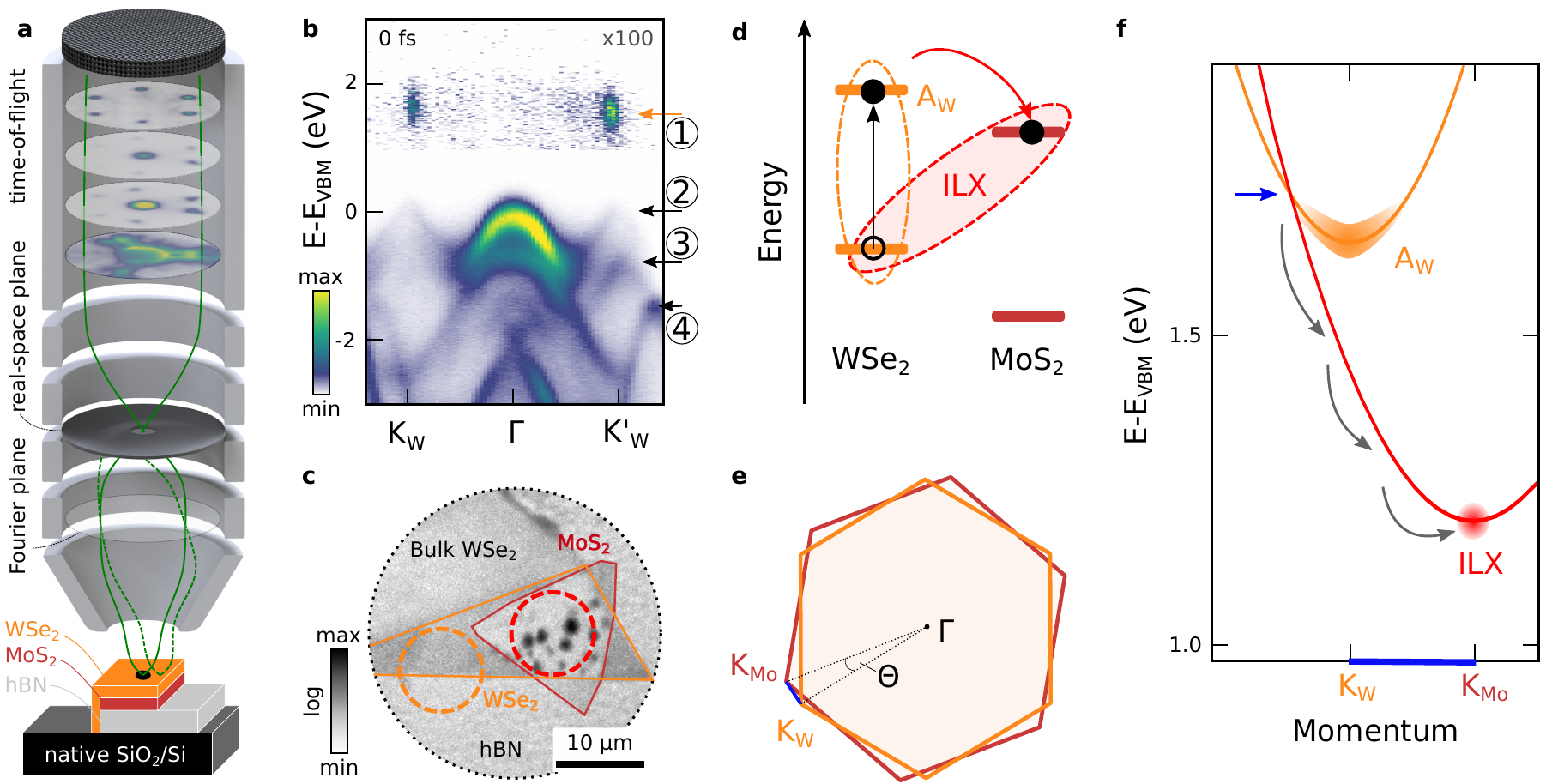}
    \caption{\textbf{Inter- and intralayer excitons in WSe$_2$/MoS$_2$ probed by femtosecond momentum microscopy.}
    \textbf{a} Illustration of the experimental setup and the WSe$_2$/MoS$_2$/hBN sample.
    \textbf{b} In a energy-momentum cut through the multidimensional photoemission data, the valence bands of (2) WSe$_2$, (3) MoS$_2$, and (4) hBN are detected. In temporal overlap of the 1.7~eV pump and the 26.5~eV probe pulses (0~fs), (1) bright A$_{\mathrm{W}}$-excitons are detected at the K$_{\mathrm{W}}$ (K'$_{\mathrm{W}}$) valleys of WSe$_2$.
    \textbf{c} The heterostructure can be identified in the real-space mode of the microscope. The regions of interest for WSe$_2$/MoS$_2$ and WSe$_2$ are indicated by red and orange circles, respectively.
    \textbf{d} Energy-level diagram of the type II band aligned heterostructure. For 1.7~eV-photons, the optical excitation occurs selectively within the WSe$_2$ layer. Subsequently, ILXs are formed where the hole and the electron contribution to the quasiparticle reside in the WSe$_2$ and MoS$_2$ layer, respectively.
    \textbf{e} The hexagonal Brillouin zones of WSe$_2$ (orange) and MoS$_2$ (brown) are misaligned by a twist angle of $\Theta=9.4\pm1.5^\circ$ (see methods and extended Fig. 4).
    \textbf{f} The excitonic band structure of the A$_{\mathrm{W}}$-exciton and the ILX (orange and red parabola). The orange shaded area represents the energy and momentum range of optically excited A$_{\mathrm{W}}$-excitons using broadband ultrashort optical pulses (h$\nu$ = 1.7 eV, $\Delta\tau \approx$50~fs, full-width-at-half-maximum $\Delta$h$\nu$ $\approx$50~meV). ILX formation can occur via scattering (not shown) and most efficiently via tunneling at the intersection of the A$_{\mathrm{W}}$-exciton and ILX paraboloids (blue arrow). The grey arrows indicate the subsequent cooling via phonon scattering~\cite{Ovesen19comphys}. The band bottom energies are extracted from experiment and the effective masses from Ref.~\onlinecite{RuizTijerina20prb}.
    }
\end{figure*}

Experimental quantitative insight into the moiré-modulated ILX formation process is, however, so far strongly limited. All-optical spectroscopy techniques are only sensitive to transitions within the light cone~\cite{Yu15prl} and thus lack the momentum information that is necessary to reconstruct the real-space distribution of the exciton wavefunction~\cite{Puschnig09sci}. Here, we overcome this experimental limitation by exploiting the full strength of multidimensional time- and angle-resolved photoelectron spectroscopy (trARPES), which is directly sensitive to the time-dependent energy-momentum-fingerprints of the probed quasiparticles~\cite{Sobota21rmp, Madeo20sci, Wallauer21nanolett, Dong20naturalsciences}. From the analysis of our trARPES data on a 9.4$\pm$1.5$^\circ$ twisted tungsten diselenide / molybdenum disulfide heterostructure (WSe$_2$/MoS$_2$), we find that ILX are formed via interlayer tunneling at the K valleys on the sub-50~fs timescale and subsequently via phonon scattering from intermediate dark excitons. In addition, we retrieve for the first time in experiment how the excitonic band structure and the real-space wavefunction is modulated by the moiré superlattice.

%%%%%%%%%%%%%%%%%%%%%%%%%%%%%%%%%%%%%%%%%%%%%%
%%%%%%%%%%%%% the trick of the experiment and static characterization of the sample
%%%%%%%%%%%%%%%%%%%%%%%%%%%%%%%%%%%%%%%%%%%%%%

We focus our study on the prototypical model system WSe$_2$/MoS$_2$ that is fabricated via exfoliation techniques and stamped onto isolating layers of hexagonal boron nitride (hBN)~\cite{Taniguchi07jcg} (see methods, extended Fig.~1 and 2). In this way, highest quality TMD heterostructures with an area of 100~$\mu m^2$ can be realized. In order to investigate these micron-scale samples, we make use of our customized time-resolved photoemission system that combines a time-of-flight momentum microscope~\cite{medjanik_direct_2017} with a high-repetition-rate high-harmonic generation beamline (Fig.~1a and methods)~\cite{Keunecke20timeresolved, Keunecke20prb}. We induce the ultrafast exciton dynamics by resonantly exciting the A$_{\mathrm{W}}$-exciton in WSe$_2$ (Fig.~1d, 1.7~eV photons). Importantly, this photon energy lies below the A$_M$-exciton resonance of MoS$_2$, which implies that all photoemission signatures of excitonic features from MoS$_2$ are built up via interlayer charge transfer from WSe$_2$ to MoS$_2$, i.e., due to the formation of ILXs. In the trARPES experiment, the electron contribution of the exciton is then probed with a time-delayed 26.5~eV pulse. The 100~$\mu$m$^2$ heterobilayer region can be identified in the real-space distribution of the measured photoelectron yield and the optical microscope image (Fig.~1c and extended Fig.~2). By placing an aperture into the real-space image plane of the momentum microscope (red circle in Fig. 1c), we can selectively probe the occupied band structure of the WSe$_2$/MoS$_2$ stack. The excellent sample quality is evidenced by the sharp spectral features of the occupied electronic structure and the signature of interlayer hybridization of the valence bands of WSe$_2$ and MoS$_2$ at the $\Gamma$ valley~\cite{Wilson17sciadv} (Fig.~1b and extended Fig.~3). In the following, we will first analyze selected time-resolved photoelectron spectroscopy data and elucidate the ILX formation mechanism. Thereafter, we make use of the momentum-resolved data collection scheme and quantify the ILX wavefunction modulation in the moiré potential.

\begin{figure}[hbt!]
    \centering
    \includegraphics[width=\linewidth]{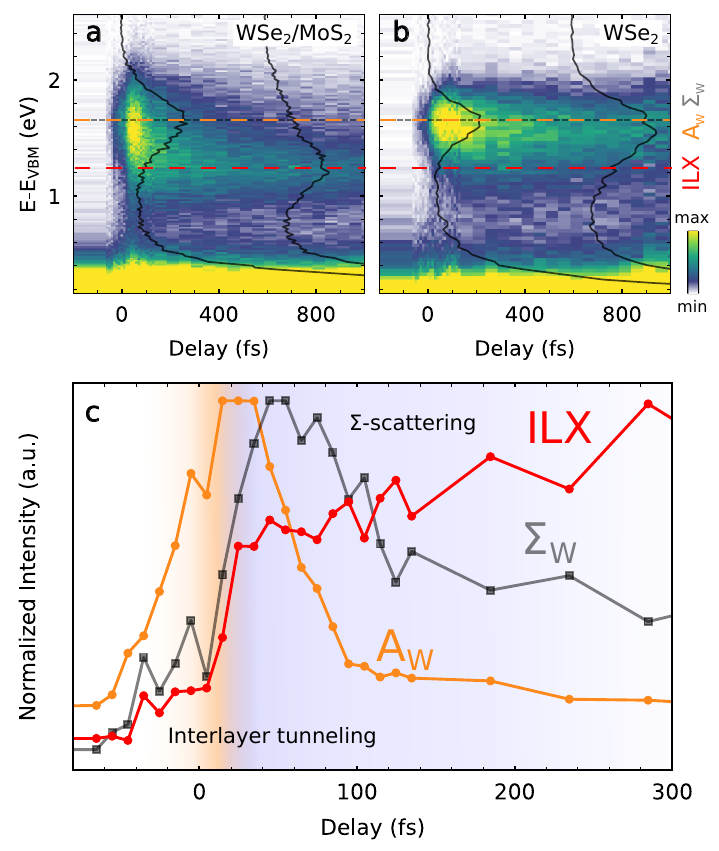}
    \caption{\textbf{Ultrafast formation dynamics of the moiré interlayer excitons.}
    \textbf{a, b} Pump-probe delay-dependent evolution of the momentum-integrated energy distribution curves for \textbf{a} WSe$_2$/MoS$_2$ and \textbf{b} WSe$_2$. In both cases, the exciton dynamics are optically induced via the excitation of the bright A$_{\mathrm{W}}$-exciton in WSe$_2$. In the monolayer, the signal decays on the picosecond timescale without a significant change in binding energy (dashed orange and grey line). In the heterobilayer, the type II band alignment facilitates the  formation of the ILXs, as is evident by the shift of spectral yield to smaller binding energies on the sub-100-fs timescale (dashed red line). The black line profiles are exemplary energy distribution curves taken at -5~fs and 585~fs.
    \textbf{c} The mechanism of the ultrafast ILX formation can be probed by visualizing the delay-dependent photoemission yield separately for the ILX, the bright A$_{\mathrm{W}}$-, and the dark $\Sigma_{\mathrm{W}}$-excitons (details on data analysis in extended Fig.~5 and 6). The two distinct charge transfer channels, i.e., interlayer tunneling and $\Sigma$-scattering, are indicated by orange and blue shading, respectively. 
    }
\end{figure}

%%%%%%%%%%%%%%%%%%%%%%%%%%%%%%%%%%%%%%%%%%%%%%
%%%%%%%%%%%%% ILX identification
%%%%%%%%%%%%%%%%%%%%%%%%%%%%%%%%%%%%%%%%%%%%%%

For type II band-aligned WSe$_2$/MoS$_2$ excited at the WSe$_2$ A$_{\mathrm{W}}$-exciton resonance, it is proposed that the ILX formation occurs via the transfer of the exciton's electron into the MoS$_2$ layer (Fig.~1d)~\cite{Rivera18natnano, Jin18natnano, Jiang21light}. At 0~fs delay between the optical pump (1.7~eV) and the extreme ultraviolet probe (26.5~eV) pulses, we explore the excitonic state in the form of the optically bright A$_{\mathrm{W}}$-exciton peak of WSe$_2$ that is detected at 1.7~eV above the valence band maximum (marked with an orange dashed line in Fig.~2a, exciton density: 3$\times$10$^{13}$cm$^{-2}$), which is in agreement with photoluminescence~\cite{Karni19prl} and earlier trARPES~\cite{Madeo20sci} experiments on the monolayer WSe$_2$. In addition, however, a second peak is formed at lower energy (red dashed line) that becomes the dominant signature on the few hundred femtosecond timescale. We identify this peak as the ILX: The photoemission signature is detected below the A$_{\mathrm{W}}$-exciton resonance at about 1.2~eV above the valence band maximum of WSe$_2$, in agreement with static photoluminescence experiments on a WSe$_2$/MoS$_2$ heterobilayer~\cite{Karni19prl}. For direct comparison, we repeated the same analysis with data obtained from the monolayer WSe$_2$ region of our sample (Fig.~2b). In this case, no spectral weight is observed in the ILX´s energetic region, which unambiguously shows that the spectral weight in the heterobilayer measurement results from the charge transfer of the electron contribution of the exciton into the MoS$_2$ layer.

%%%%%%%%%%%%%%%%%%%%%%%%%%%%%%%%%%%%%%%%%%%%%%
%%%%%%%%%%%%% ILX formation
%%%%%%%%%%%%%%%%%%%%%%%%%%%%%%%%%%%%%%%%%%%%%%

The exact mechanism of the ILX formation and the according ultrafast charge separation is still a major open question~\cite{Jin18natnano}. It has been proposed that the ILXs can be formed via interlayer tunneling of its electron contribution~\cite{Ovesen19comphys, Merkl19natmat}, or, alternatively, via the intermediate formation of dark intralayer excitons, where the electron contribution is first scattered to the $\Sigma$ or $\Gamma$ valleys and, subsequently, transferred to the neighbouring layer~\cite{Wang17prb, Kunstmann18natphys,Wallauer20prb}. Our experiment, here, shows very striking signatures in the delay-dependent transfer of spectral weight between the bright WSe$_2$ A$_{\mathrm{W}}$-exciton, the dark WSe$_2$ $\Sigma_{\mathrm{W}}$-exciton, and the ILX (Fig.~2c, details on data handling in extended Fig.~5 and 6). Initially, during the timescale of the pump excitation (orange shaded area in Fig.~2c), we find that the build-up of dark WSe$_2$ $\Sigma_{\mathrm{W}}$-excitons and ILXs is most efficient (black and red in Fig.~2c, respectively). However, for $\approx$30~fs, coinciding with the maximum yield of the A$_{\mathrm{W}}$-exciton (orange), the efficiency of the ILX formation process is drastically diminished. Subsequently, for times larger than $\approx$50 fs, the ILX build-up then resembles the decay of spectral weight of the WSe$_2$ $\Sigma_{\mathrm{W}}$-exciton (blue shaded area in Fig.~2c). Therefore, our data indicates that ILX build-up occurs dominantly via intermediate formation of dark intralayer excitons and subsequent transfer to the neighboring layer, which is in agreement with Refs.~\onlinecite{Wang17prb,Kunstmann18natphys,Wallauer20prb}. However, on the short timescale, and in particular for delays <50~fs, this process is clearly too slow to explain the initial and surprisingly steep rise of ILX spectral weight. This observation can be explained, in addition, via resonant tunneling at the intersection points of the inter- and intralayer dispersion relations (Fig.~1f), where both energy and the in-plane momentum can simultaneously be conserved~\cite{Merkl19natmat,Ovesen19comphys}. Such resonant tunneling processes indeed become possible due to the broadband optical excitation: A$_{\mathrm{W}}$-excitons are not only excited at the very bottom of the excitonic band, but also at higher energies and finite momenta reaching up to the intersection points of the exciton dispersion relations (orange shaded area in Fig.~1f). Consequently, during the optical excitation, ILX formation via resonant tunneling is possible and can explain the steep rise of the ILX formation, which coincides with the rise of the A$_{\mathrm{W}}$-exciton spectral weight. This ILX formation channel, however, rapidly closes after the optical excitation, when no more A$_{\mathrm{W}}$-excitons at sufficiently high energies are generated and the residual A$_{\mathrm{W}}$-excitons have scattered in momentum space and also towards the bottom of the A$_{\mathrm{W}}$-exciton band. Thus, by making use of the strength of the femtosecond momentum microscopy experiment, we are able to identify the cooperative dynamics of the ILX formation process directly in the time-domain and show that both suggested ILX generation processes are operative

\begin{figure*}[hbt!]
    \centering
    \includegraphics[width=\linewidth]{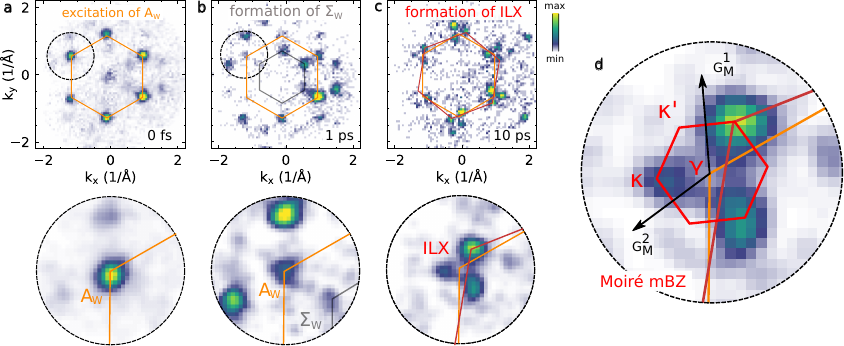}
    \caption{\textbf{Momentum-fingerprints of the moiré interlayer exciton in WSe$_2$/MoS$_2$.}
    Momentum fingerprints of \textbf{a} the bright A$_{\mathrm{W}}$-exciton, \textbf{b} the dark $\Sigma_{\mathrm{W}}$-exciton, and \textbf{c} the moiré modulated ILX. The bottom row shows zoom-ins from the circled area in the top row. The pump-probe delay of the three momentum maps are noted in the figure.
    \textbf{a} The optical excitation ($p$-polarized, 1.7~eV-photons) is carried out resonantly with the A$_{\mathrm{W}}$-exciton at the K$_{\mathrm{W}}$-valleys of the WSe$_2$ Brillouin zone (orange hexagon). Additional spectral weight in the middle of the Brillouin zone arises due the well-known laser-assisted photoelectric effect~\cite{miaja2006laser, Keunecke20prb}.
    \textbf{b} Spectral weight is transferred within the WSe$_2$ layer from the bright A$_{\mathrm{W}}$ to the dark $\Sigma_{\mathrm{W}}$-excitons (small grey hexagon). 
    \textbf{c} The momentum fingerprint of the moiré-modulated ILX is a three-peak structure centered around the K$_{\mathrm{W}}$ and K'$_{\mathrm{W}}$ valleys (orange hexagon). Notably, only one of the three signatures coincidences with the K$_{\mathrm{Mo}}$ or K'$_{\mathrm{Mo}}$ valleys (brown hexagon). 
    \textbf{d} The three-peak signature of the ILX can be constructed within the moiré mini Brillouin zone (red hexagon). The ILX photoemission signatures are found at the mBZ corners, i.e., at the $\kappa$ valleys; the mBZ center $\gamma$ lies on the former K$_{\mathrm{W}}$ valley and the moiré lattice vector G$_{\rm moire}^{1,2}$ is indicated by black arrows.
    }
\end{figure*}

%%%%%%%%%%%%%%%%%%%%%%%%%%%%%%%%%%%%%%%%%%%%%%
%%%%%%%%%%%%% moire identification
%%%%%%%%%%%%%%%%%%%%%%%%%%%%%%%%%%%%%%%%%%%%%%
The full potential of the experiment unfolds when we explicitly use the momentum-resolved data collection scheme in order to study how the ILX interacts with the moiré potential that is built up by the misalignment of the Brillouin zones of WSe$_2$ and MoS$_2$ (Fig. 1e). Specifically, we illuminate how the ILX wavefunction is localized and modulated within the moiré potential. Therefore, Fig.~3a, b, and c show the momentum fingerprints of the A$_{\mathrm{W}}$-exciton, the $\Sigma_{\mathrm{W}}$-exciton, and the ILX. While the measured momentum fingerprints of the intralayer excitons are in agreement with recent reports~\cite{Madeo20sci, Wallauer21nanolett, Dong20naturalsciences}, here, we focus on the so far unexplored momentum-resolved fingerprint of the ILX. Without consideration of the moiré superlattice, one would expect to detect photoemission yield at the in-plane momentum of the electron contribution to the ILX, i.e., at the K$_{\mathrm{Mo}}$ and K'$_{\mathrm{Mo}}$ valleys of MoS$_2$ (edges of the brown hexagon in Fig.~3c). Astonishingly, the measured momentum fingerprint shows a strikingly richer structure: Instead of a single peak at the K$_{\mathrm{Mo}}$ (K'$_{\mathrm{Mo}}$) valleys, we observe three peaks which are centered around the K$_{\mathrm{W}}$ (K'$_{\mathrm{W}}$) valleys (orange hexagon in Fig.~3c). While the momentum-integrated measurements have identified emission energies consistent with ILXs, the interaction with an underlying moiré lattice has remained opaque. Therefore, in the following, we will clarify how the measured threefold momentum fingerprint directly results from the modulation of the ILX wavefunction within the moiré potential.

%%%%%%%%%%%%%%%%%%%%%%%%%%%%%%%%%%%%%%%%%%%%%%
%%%%%%%%%%%%% moire mBZ
%%%%%%%%%%%%%%%%%%%%%%%%%%%%%%%%%%%%%%%%%%%%%%

In Fig.~3d, we overlay the moiré mini-Brillouin zone (mBZ) with the momentum-resolved photoemission data (details in the method section). Within the mBZ, one can now identify that the measured three-peak structure perfectly coincides with the geometrically constructed momenta of the three $\kappa$ ($\kappa$') valleys of the mBZ; the $\gamma$ valley of the mBZ is located on the K$_{\mathrm{W}}$ (K'$_{\mathrm{W}}$) valleys, respectively. In other words, the threefold momentum fingerprint of the ILX results from the modulation of the ILX within the moiré superlattice in real-space. A priori, it might not be directly obvious why a threefold (and not sixfold) symmetric photoemission signature is detected within the moiré mBZ. However, theoretical calculations of the excitonic band structure within the mBZ exactly indicate this symmetry because of the two-atomic basis of TMDs~\cite{Brem20nanolett, Alexeev19nat}. Here, our experiment provides first direct experimental access to the dispersion relation of the excitonic band structures and verifies these calculations.

%%%%%%%%%%%%%%%%%%%%%%%%%%%%%%%%%%%%%%%%%%%%%%
%%%%%%%%%%%%% reconstruction
%%%%%%%%%%%%%%%%%%%%%%%%%%%%%%%%%%%%%%%%%%%%%%

\begin{figure*}[hbt!]
    \centering
    \includegraphics[width=\linewidth]{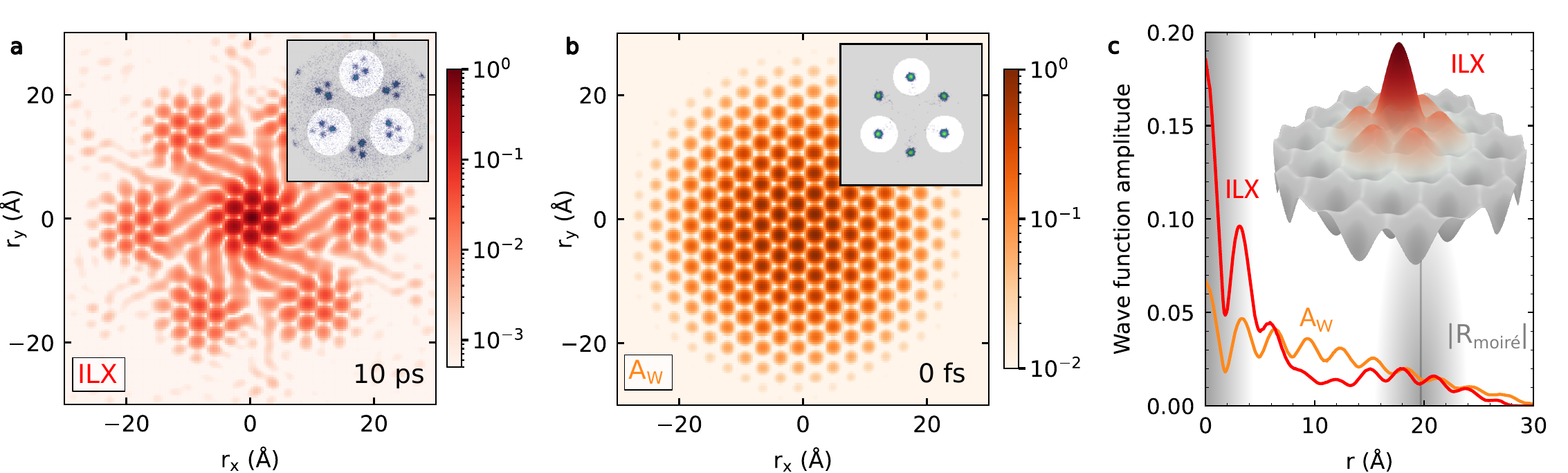}
    \caption{\textbf{Real-space modulation of the interlayer exciton wavefunction within the moiré potential.}
    Real-space reconstructions of the probability density of the \textbf{a} ILX and the \textbf{b} WSe$_2$ A$_{\mathrm{W}}$-exciton. The insets show the symmetrized data and a mask which was used to select a single spin-valley component for the reconstruction. The modulation of the probability density is composed of an isotropic decay and a fast oscillation due to the lattice potential.
    \textbf{a} In addition, the probability density of the ILX is modulated by the moiré potential,
    which is, however, \textbf{b} not observed in the case of the WSe$_2$ A$_{\mathrm{W}}$-exciton. The reconstructed WSe$_2$ A$_{\mathrm{W}}$-exciton wavefunction is in qualitative agreement with the reconstruction reported for monolayer WSe$_2$~\cite{Man21sciadv}.
    \textbf{c} The wave function amplitude of the ILX (red line) is maximal at the position of the moiré potential minima, i.e., at $r=0$ and $\left|R_{\rm moire}\right|=1.96\pm0.28$~nm (grey shading). The Bohr radius of the ILX (red line) and the WSe$_2$ A$_{\mathrm{W}}$-exciton (orange line) can be extracted to 1~nm and 1.3~nm, respectively (root-mean-square). The inset schematically shows how the ILX wavefunction (red) is modulated in the moiré potential (grey). Data set \textbf{a} and \textbf{b} were obtained under $p$- and $s$-polarized excitation, respectively.
    }
\end{figure*}

Having identified the threefold momentum structure as the consequence of the real-space modulation of the ILX wavefunction by the moiré potential, we now reconstruct the real-space electron contribution to ILX wavefunction using photoemission orbital tomography~\cite{Puschnig09sci}. The excitonic wavefunction has been intensely discussed in theoretical works~\cite{Hongyi17sciadv, Wu18prb, Seyler19nat, Alexeev19nat, Tran19nat}, however, until this day, it has remained experimentally inaccessible. We employ the the relation $I(k_x,k_y)\propto \left|FT \{ \Psi\left(r_x,r_y \right) \}\right|^2$ that connects the real-space wavefunction $\Psi\left(r_x,r_y \right)$ with the momentum-resolved photoemission intensity $I(k_x,k_y)$ within the plane wave approximation (see methods)~\cite{Puschnig09sci, jansen_efficient_2020}. In Fig.~4a, we show the real-space probability density of the ILX. Unambiguously, we observe that the ILX is strongly affected by the moiré potential, which results in a significant modulation of the ILX wavefunction with the period of the moiré lattice constant ($\left|R_{\rm moire }\right|$ in Fig. 4c). Intriguingly, our reconstruction further quantifies that for the WSe$_2$/MoS$_2$ heterobilayer with a 9.4$\pm$1.5$^\circ$ twist angle, approximately 60$\%$ of the probability density reside within one moiré unit cell. The six neighbouring potential minima contain approximately 6$\%$ each (Fig.~4a and 4c). From this data, we can extract the exciton Bohr radius of the ILX to $\approx 1$~nm. Furthermore, in Fig.~4b/c, we show that the wavefunction of the intralayer A$_{\mathrm{W}}$-exciton, in contrast, is not modulated within the moiré potential. This observation is a direct experimental verification that the charge separation between both layers, i.e., the formation of ILXs, is a key step towards the realization of new phases where moiré and exciton physics is correlated.

%%%%%%%%%%%%%%%%%%%%%%%%%%%%%%%%%%%%%%%%%%%%%%
%%%%%%%%%%%%% conclusion, and what's next?
%%%%%%%%%%%%%%%%%%%%%%%%%%%%%%%%%%%%%%%%%%%%%%

In summary, we have employed the newly developed strength of femtosecond momentum microscopy in order to quantify the formation dynamics of ILXs in a twisted WSe$_2$/MoS$_2$ heterostructure. Specifically, we elucidate that ILXs are most efficiently formed via interlayer tunneling at the K valleys on the sub-50~fs timescale and subsequently via scattering from intermediate dark exciton levels. Even more, with the unprecedented experimental detection of the ILX momentum fingerprint, we reconstruct that the ILX wavefunction is significantly modulated within the moiré potential. Our experimental approach is very general and not limited to the specific system discussed here, but can be applied to arbitrary stacked, twisted, or strained van-der-Waals heterostructures. Our results thus open highest-detail quantitative insight into the properties of exotic quasiparticles in quantum materials and thus lays the foundation for future in depth exploration of correlated moiré and exciton physics on the nanometer length- and femtosecond time-scales.

%%%%%%%%%%%%%%%%%%%%%%%%%%%%%%%%%%%%%%%%%%%%%%
%%%%%%%%%%%%% acknowledgment  // Author contribution and so on
%%%%%%%%%%%%%%%%%%%%%%%%%%%%%%%%%%%%%%%%%%%%%%

\section{ACKNOWLEDGEMENTS}

This work was funded by the Deutsche Forschungsgemeinschaft (DFG, German Research Foundation) - 432680300/SFB 1456, project B01 and 217133147/SFB 1073, projects B07 and B10. G.S.M.J. acknowledges financial support by the Alexander von Humboldt Foundation. A.A. and S.H. acknowledge funding from EPSRC (EP/T001038/1, EP/P005152/1). A.A. acknowledges financial support by the Saudi Arabian Ministry of Higher Education. K.W. and T.T. acknowledge support from the Elemental Strategy Initiative
conducted by the MEXT, Japan (Grant Number JPMXP0112101001) and JSPS
KAKENHI (Grant Numbers 19H05790, 20H00354 and 21H05233).

\section{AUTHOR CONTRIBUTIONS}
D.St., R.T.W., S.S., G.S.M.J., S.H., M.R. and S.M. conceived the research. D.Sch., J.P.B. and W.B. carried out the time-resolved momentum microscopy experiments. D.Sch. and J.P.B. analyzed the data. W.B., D.R.L., and G.S.M.J. carried out the real-space reconstruction of the momentum fingerprints. A.A. fabricated the samples. All authors discussed the results. M.R. and S.M. were responsible for the overall project direction and wrote the manuscript with contributions from all co-authors. K.W. and T.T. synthesized the hBN crystals.

%\section{REFERENCES}
%\section{REFERENCES}
%merlin.mbs apsrev4-1.bst 2010-07-25 4.21a (PWD, AO, DPC) hacked
%Control: key (0)
%Control: author (8) initials jnrlst
%Control: editor formatted (1) identically to author
%Control: production of article title (-1) disabled
%Control: page (0) single
%Control: year (1) truncated
%Control: production of eprint (0) enabled
%\begin{thebibliography}{41}%

%merlin.mbs apsrev4-1.bst 2010-07-25 4.21a (PWD, AO, DPC) hacked
%Control: key (0)
%Control: author (0) dotless jnrlst
%Control: editor formatted (1) identically to author
%Control: production of article title (0) allowed
%Control: page (1) range
%Control: year (0) verbatim
%Control: production of eprint (0) enabled
%

%\end{thebibliography}%

%\end{document}

%\bibliographystyle{unsrt}
%\bibliographystyle{abbrv}
%\bibliography{bibtexfile}% Produces the bibliography via BibTeX.
%\printbibliography

%%%%%%%%%%%%%%%%%%%%%%%%%%%%%%%%%%%%%%%%%%%%%%%%%%%%%%%%%%%%%
%%%%%%%%%%%%%%%%%%%%%%%%%%%%%%%%%%%%%%%%%%%%%%%%%%%%%%%%%%%%%%%
%%%%%%%%%%%%%%%%%%%%%%%%%%%%%%%%%%%%%%%%%%%%%%%%%%%%%%%%%%%
%%%%%%%%%%%%%%%%%%%%%%%%%%%%%%%%%%%%%%%%%%%%%%%%%%%%%%%%%%

%\newpage

%\section{Supplemental material}

\end{document}

%\preprint{AIP/123-QED}

\title[Formation of moiré interlayer excitons in space and time]{Formation of moiré interlayer excitons in space and time}

% Force line breaks with \\

\author{David Schmitt} %
\address{I. Physikalisches Institut, Georg-August-Universit\"at G\"ottingen, Friedrich-Hund-Platz 1, 37077 G\"ottingen, Germany}

\author{Jan Philipp Bange} %
\address{I. Physikalisches Institut, Georg-August-Universit\"at G\"ottingen, Friedrich-Hund-Platz 1, 37077 G\"ottingen, Germany}

\author{Wiebke Bennecke} %
\address{I. Physikalisches Institut, Georg-August-Universit\"at G\"ottingen, Friedrich-Hund-Platz 1, 37077 G\"ottingen, Germany}

\author{AbdulAziz AlMutairi} 
\address{Department of Engineering, University of Cambridge, Cambridge CB3 0FA, U.K.}

\author{Kenji Watanabe} %
\address{Research Center for Functional Materials, National Institute for Materials Science, 1-1 Namiki, Tsukuba 305-0044, Japan}

\author{Takashi Taniguchi} %
\address{International Center for Materials Nanoarchitectonics, National Institute for Materials Science, 1-1 Namiki, Tsukuba 305-0044, Japan}

\author{Daniel Steil} %
\address{I. Physikalisches Institut, Georg-August-Universit\"at G\"ottingen, Friedrich-Hund-Platz 1, 37077 G\"ottingen, Germany}

\author{D. Russell Luke} %
\address{Institute for Numerical and Applied Mathematics, Georg-August-Universit\"at G\"ottingen, Lotzestrasse 16-18, 37083 G\"ottingen, Germany}

\author{R. Thomas Weitz} %
\address{I. Physikalisches Institut, Georg-August-Universit\"at G\"ottingen, Friedrich-Hund-Platz 1, 37077 G\"ottingen, Germany}

\author{Sabine Steil} 
\address{I. Physikalisches Institut, Georg-August-Universit\"at G\"ottingen, Friedrich-Hund-Platz 1, 37077 G\"ottingen, Germany}

\author{G.~S.~Matthijs~Jansen} %
\address{I. Physikalisches Institut, Georg-August-Universit\"at G\"ottingen, Friedrich-Hund-Platz 1, 37077 G\"ottingen, Germany}

\author{Stephan Hofmann} 
\address{Department of Engineering, University of Cambridge, Cambridge CB3 0FA, U.K.}

\author{Marcel Reutzel} \email{marcel.reutzel@phys.uni-goettingen.de}%
\address{I. Physikalisches Institut, Georg-August-Universit\"at G\"ottingen, Friedrich-Hund-Platz 1, 37077 G\"ottingen, Germany}

\author{Stefan Mathias} \email{smathias@uni-goettingen.de}%
\address{I. Physikalisches Institut, Georg-August-Universit\"at G\"ottingen, Friedrich-Hund-Platz 1, 37077 G\"ottingen, Germany}

\begin{figure*}[hbt!]
    \centering
    \includegraphics[width=\linewidth]{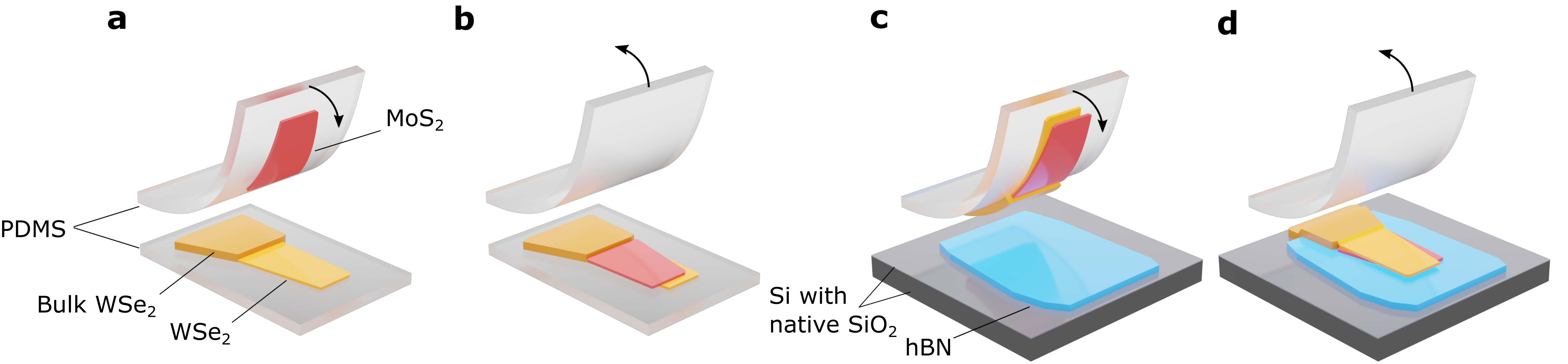}
    \caption{\textbf{Fabrication of the WSe$_2$/MoS$_2$/hBN heterostructure.}
    \textbf{a-d} Schematic of the WSe$_2$/MoS$_2$/hBN heterostructure assembly on a Si substrate with a native oxide layer. 
    \textbf{a} A PDMS stamp with monolayer MoS$_2$ flake (red) is aligned with another PDMS stamp holding a monolayer WSe$_2$ (yellow), then the top PDMS stamp is brought in contact with bottom PDMS. 
    \textbf{b} The top PDMS stamp is then withdrawn, leaving the MoS$_2$ on top of the WSe$_2$.
    \textbf{c} The PDMS stamp holding the WSe$_2$/MoS$_2$ is then brought into contact with SiO$_2$/Si substrate with hBN (blue). d the PDMS is then withdrawn, leaving behind the final WSe$_2$/MoS$_2$/hBN heterostructure. }
    \label{render-howto}
\end{figure*}

\begin{figure*}[hbt!]
    \centering
    \includegraphics[width=\linewidth]{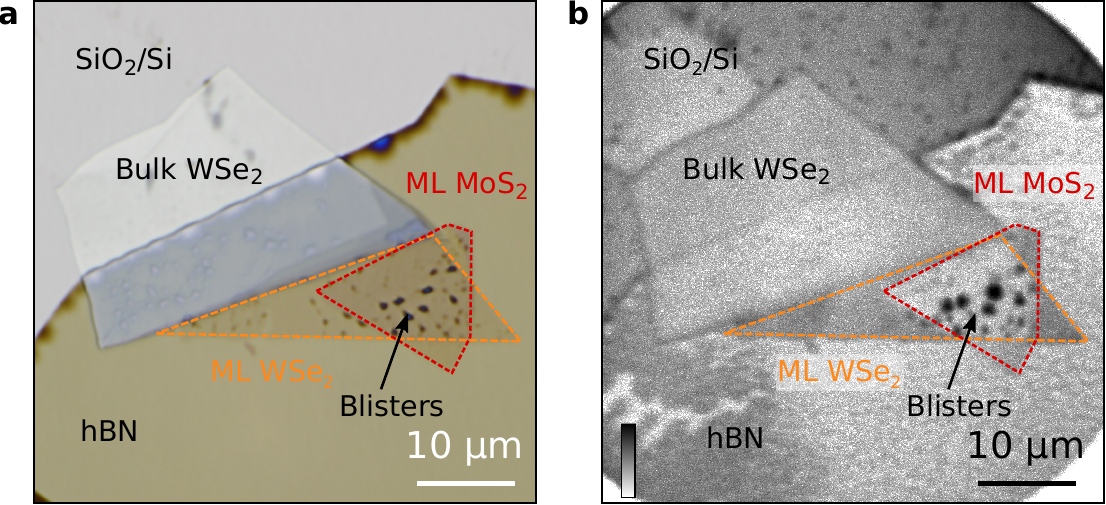}
    \caption{\textbf{Real-space characterization of the WSe$_2$/MoS$_2$/hBN heterostructure.} 
    The WSe$_2$/MoS$_2$/hBN heterostructure, the uncovered SiO$_2$/Si substrate, bulk hBN, the WSe$_2$ and MoS$_2$ monolayers, and bulk WSe$_2$ are labelled in the \textbf{a} optical microscope and the \textbf{b} photoemission real-space image. Point-like structures (blisters) in the heterostructure region can be attributed to residual gas trapped either at the MoS$_2$/hBN or the WSe$_2$/MoS$_2$ interface. Photoelectrons in the real-space image are excited by 4.96~eV photons from a UV-diode.}
    \label{optical_vs_PEEM}
\end{figure*}

\noindent\textbf{Heterostructure fabrication.} The WSe$_2$/MoS$_2$/hBN heterostructure was assembled using mechanical exfoliation and dry transfer,  as summarized in extended Fig.~\ref{render-howto}. First, a  p$^+$-Si substrate (1-10~$\Omega$cm) with polished native oxide was plasma cleaned with O$_2$ gas (100~W, 10~sccm, 10~mins). hBN was immediately mechanically exfoliated on the Si substrate using standard office tape. Using optical contrast, an hBN flake with thickness between 20-30~nm was identified. In parallel, MoS$_2$ and WSe$_2$ (HQ graphene) were mechanically exfoliated using blue tape (Ultron Systems, Inc. 1008R-6.0) on a Polydimethylsiloxane (PDMS) sheet (Gel-Pak PF-20/17-X4). Similar to hBN exfoliation, optical contrast was used to identify monolayer MoS$_2$ and WSe$_2$. Unlike the standard dry transfer assembly, the assembly of the heterostructure was started by assembling the WSe$_2$/MoS$_2$ heterostructure on PDMS first. The two flakes were aligned visually and the MoS$_2$ flake was dry transferred on top of the WSe$_2$ flake on PDMS. Then the WSe$_2$/MoS$_2$ heterostructure was dry transferred on top of the hBN. During the last transfer, the multilayer part of the WSe$_2$ flake was intentionally placed in direct contact with the Si substrate to reduce sample charging (cf.~extended Fig.~\ref{optical_vs_PEEM}a). It is worth noting that while the exfoliation part was done in ambient condition, the heterostructure assembly was done in a controlled Ar environment in a glovebox with O$_2$ and H$_2$O <0.1 ppm. An optical microscope image of the van-der-Waals stack is shown in extended Fig.~\ref{optical_vs_PEEM}a.

\vspace{1cm}

\noindent\textbf{Sample preparation for photoemission spectroscopy.} The silicon wafer with the heterostructure is clamped onto a sample holder under ambient conditions and transferred into ultrahigh vacuum (UHV). In an UHV preparation chamber (background pressure < 5$\times$10$^{-9}$~mbar), the sample is annealed for 2~hours at a temperature of approximately 400$^\circ$C. Subsequently, the sample is transferred into the momentum microscope. All experiments have been performed at room temperature at a background pressure of approximately 1$\times$10$^{-10}$~mbar. 

\vspace{1cm}

\noindent\textbf{Femtosecond momentum microscopy.} The experimental setup is detailed and benchmarked in Ref.~\onlinecite{Keunecke20timeresolved}. It consists of two major parts, namely, (i) a time-of-flight momentum microscope~\cite{medjanik_direct_2017} (METIS, Surface Concepts), shown schematically in Fig.~1a of the main text, and (ii) a high-power femtosecond laser system (Active Fiber Systems). In the following, we will briefly describe the experimental setup.

\begin{figure*}[hbt!]
    \centering
    \includegraphics[width=\linewidth]{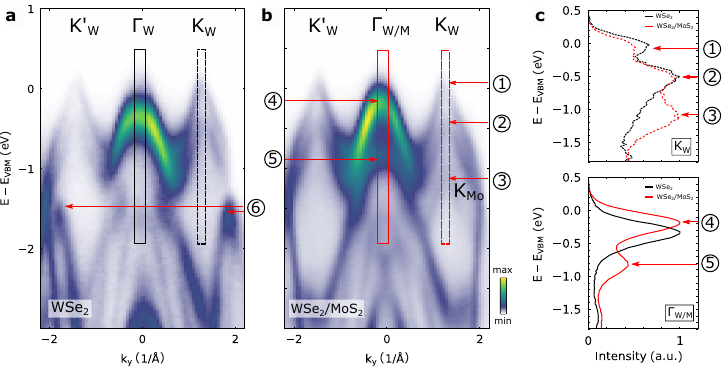}
    \caption{\textbf{Static band mapping of the monolayer WSe$_2$ and the heterobilayer WSe$_2$/MoS$_2$.}
    \textbf{a,b} Energy-momentum representation of the static photoemission intensity obtained in the momentum microscopy experiment. The important spectroscopic features are labelled in the figure: (1, 2) spin-split valence bands of WSe$_2$; (3) valence band of MoS$_2$; (4, 5) valence bands at the $\Gamma_{\rm W,Mo}$ valley; (6) valence band of hBN.
    \textbf{c, d} Energy distribution curves taken around the K$_{\mathrm{W}}$ and $\Gamma_{\rm W,Mo}$ ($\Gamma_{\mathrm{W}}$) valley.}
    \label{mono_vs_bilage}
\end{figure*}

(i) The strength of the time-of-flight momentum microscope is the simultaneous measurement of the 2D in-plane momenta and the kinetic energy of the photoelectrons within the full photoemission horizon~\cite{medjanik_direct_2017}. In order to study high-quality TMD heterostructures with a diameter of approximately 10$\mu$m, however, the key advantage of the setup lies in the microscopy-type photoelectron detection scheme: In the electrostatic lens system, a real- and a reciprocal-space image of the photoelectrons is formed and either image can be projected onto the photoelectron detector. The real-space mode of the microscope is used to map the spatial structure of the sample system (Fig.~1c, extended Fig.~\ref{optical_vs_PEEM}b). In direct comparison to an optical microscope image (extended Fig.~\ref{optical_vs_PEEM}a), the monolayer WSe$_2$ and MoS$_2$ regions as well as the WSe$_2$/MoS$_2$ heterobilayer region can be identified via the differing photoemission contrast. By placing an aperture into the real-space plane of the microscope, either photoelectrons originating from the WSe$_2$ mono- or the WSe$_2$/MoS$_2$ heterobilayer can be selected and projected onto the detector (extended Fig.~\ref{mono_vs_bilage}). By exploiting this capability of the time-of-flight momentum microscope, we collect in-plane momentum- and energy-resolved photoelectron distributions for the regions of interest highlighted by circles in Fig. 1c of the main text.

(ii) The laser setup is based on a 300~W fiber laser system (Active Fiber Systems) that operates at a repetition rate of 500~kHz and drives a high-harmonic generation beamline and a high-power optical parametric amplifier (OPA, Orpheus-F/HP from Light Conversion). In order to first induce and subsequently probe excitonic dynamics occurring in the TMD heterostructure, we employ a pump-probe scheme: First, bright WSe$_2$ A$_{\mathrm{W}}$ excitons are excited with light pulses generated with the OPA (1.7~eV, $s$-polarized (if not otherwise stated), 50~fs, 200~µJ/cm$^2$). Subsequently, the femto- to picosecond evolution of the intra- and interlayer excitonic dynamics is probed with an extreme-ultraviolet light pulse (26.5~eV, 20~fs), which photoemits the electron contribution of the quasiparticle into the detector.

\vspace{1cm}

\noindent\textbf{Real-space imaging and static band mapping of WSe$_2$/MoS$_2$/hBN.} After preparation of the van-der-Waals stack for the momentum microscopy experiment, we first perform real-space imaging of the sample with a continuous wave UV-diode delivering 4.96~eV photons. In extended Fig.~\ref{optical_vs_PEEM}, the photoemission real-space map is compared with an optical microscope image. In both images, the WSe$_2$/MoS$_2$/hBN heterostructure, the doped Si substrate, the bulk hBN, the WSe$_2$ and MoS$_2$ monolayers, and the bulk WSe$_2$ can be distinguished. 

Having identified the regions of interested, we place an aperture into the real-space image of the microscope in order to selectively probe the energy- and momentum-resolved photoelectron distribution of the WSe$_2$ monolayer and the WSe$_2$/MoS$_2$ heterobilayer (circles in Fig. 1c); static band mapping of the occupied electronic band structure is shown in extended Fig.~\ref{mono_vs_bilage}. The high quality of the van-der-Waals stack is evident by the sharp features in the band structure and, in particular, by the well-resolved spin-splitting of the WSe$_2$ valence bands at the K$_{\mathrm{W}}$ and K'$_{\mathrm{W}}$ valleys~\cite{Ulstrup19apl} [(1) and (2) extended Fig.~\ref{mono_vs_bilage}]. In addition, only in the heterobilayer region, we resolve clear signatures of the valence band maximum of MoS$_2$ at -1.1~eV with respect to the valence band maximum of WSe$_2$ [(3) in extended Fig.~\ref{mono_vs_bilage}b]. Due to interlayer interaction between the WSe$_2$ and MoS$_2$ layers, we clearly resolve the expected hybridized valence bands at the $\Gamma_{W,M}$ valley [(4) and (5) in extended Fig.~\ref{mono_vs_bilage}b]~\cite{Wilson17sciadv}. The observation of these hybridized bands is a clear signature that the blisters found in the real-space image in extended Fig.~\ref{optical_vs_PEEM} do not dominate the photoemission yield from the heterobilayer. In contrast, in the monolayer WSe$_2$ region, the valence band at the $\Gamma_{\mathrm{W}}$ valley is a single band (extended Fig.~\ref{mono_vs_bilage}a). Furthermore, the valence band maximum is localized at the  K$_{\mathrm{W}}$ (K'$_{\mathrm{W}}$) valley, as expected for the monolayer limit of WSe$_2$ where it becomes a direct band gap semiconductor~\cite{Wilson17sciadv}. Furthermore, we observe a clear signature of the valence band of hBN that we label with (6) in extended Fig.~\ref{mono_vs_bilage}.

\vspace{1cm}

\noindent\textbf{Twist-angle determination of the WSe$_2$/MoS$_2$ stack.} We quantify the twist angle of the WSe$_2$/MoS$_2$ heterostructure by evaluating (i) the momentum positions of the K and M valleys of WSe$_2$ and MoS$_2$ in the occupied electronic band structure and by (ii) analyzing the three-peak momentum fingerprint of the ILX. 

\begin{figure*}[hbt!]
    \centering
    \includegraphics[width=\linewidth]{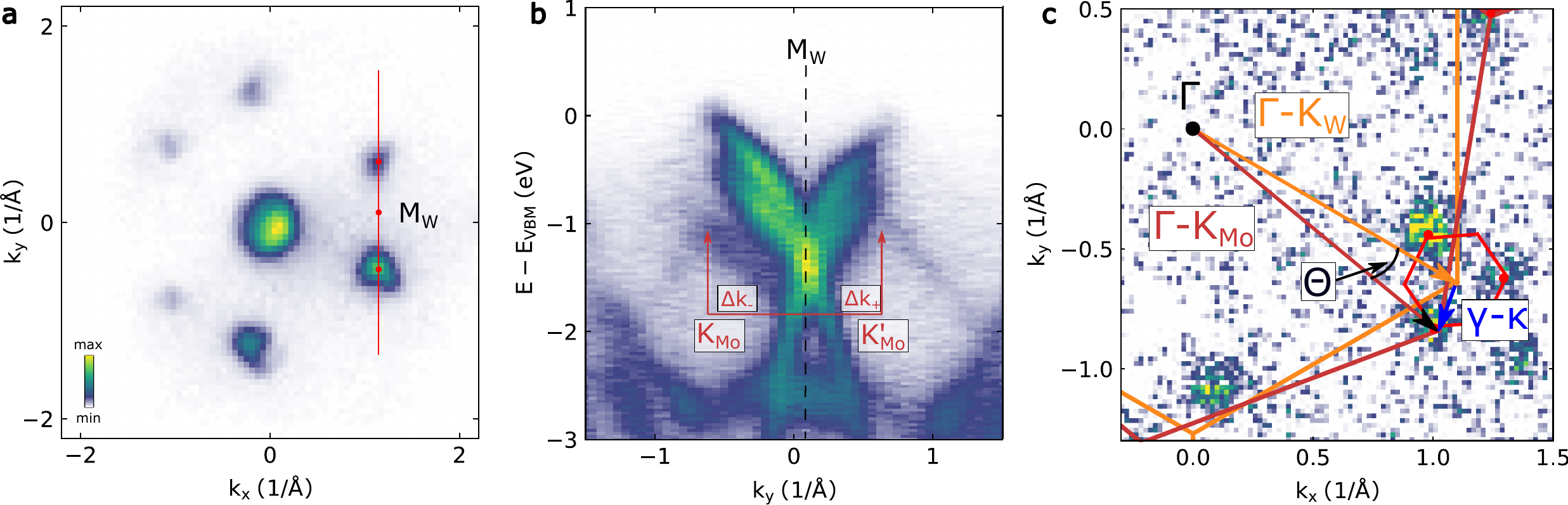}
    \caption{\textbf{Estimation of the twist angle of the heterostack in the static photoemission experiment.}
    \textbf{a} The momentum-map shows a cut at the valence band maximum of WSe$_2$. The red line indicates the momentum-direction for which the energy momentum cut is shown in \textbf{b}.
    \textbf{b} Energy-momentum cut along the K'$_{\mathrm{W}}$-M$_{\mathrm{W}}$-K$_{\mathrm{W}}$ direction. The band maxima at K$_{\mathrm{Mo}}$ and K'$_{\mathrm{Mo}}$ of the valence band of MoS$_2$ are indicated with brown arrows, and the M$_{\mathrm{W}}$ point of WSe$_2$ is indicated by a black dashed line. The twist angle can be estimated by the difference of the vectors $\Delta$k$_{-}$ and $\Delta$k$_{+}$ that describe the distance of the K$_{\mathrm{Mo}}$ (K'$_{\mathrm{Mo}}$) valleys from the M$_{\mathrm{W}}$ point.
    \textbf{c} The twist-angle $\Theta$ can be further quantified with much higher accuracy from the three-fold momentum signature of the ILX by evaluating the angle between the $\Gamma$-K$_{\mathrm{W}}$ and $\Gamma$-K$_{\rm Mo}$ vectors.
    }
    \label{twist_static_bandstructure}
\end{figure*}

(i) The angle-resolved photoemission experiment intrinsically contains information on the twist angle of WSe$_2$ and MoS$_2$; the data shown in extended Fig.~\ref{twist_static_bandstructure} gives a first estimate of the respective alignment of the Brillouin zones. In the energy-momentum cut along the K'$_{\mathrm{W}}$-M$_{\mathrm{W}}$-K$_{\mathrm{W}}$ direction (see Fig.~\ref{twist_static_bandstructure}a) in extended Fig.~\ref{twist_static_bandstructure}b, the M$_{\mathrm{W}}$ point of WSe$_2$ and the valence band maxima of MoS$_2$ are indicated by a black dashed line and brown arrows (K$_{\mathrm{Mo}}$ and K'$_{\mathrm{Mo}}$), respectively. We compare the relative distance between the K$_{\mathrm{Mo}}$ (K'$_{\mathrm{Mo}}$) valleys and the M$_{\mathrm{W}}$ point and find that $\Delta$k$_+$<$\Delta$k$_-$. From this analysis, we find that the twist angle of both layers must be in the range of roughly 5$^\circ$ to 15$^\circ$.%, as, in the case of a perfectly aligned heterostructure,  $\Delta$k$_+$ would be equal to  $\Delta$k$_-$. 
    
(ii) In addition, we quantify the twist angle with much improved accuracy by analyzing the momentum fingerprint of the ILX and use the fact that the photoemission yield of the ILX coincides with the high symmetry points of the moiré mini Brillouin zone. In extended Fig.~\ref{twist_static_bandstructure}c, we fit the momentum distribution of the three-fold momentum-fingerprint with three two-dimensional Gaussian functions (not shown) and subsequently plot the lattice vectors $\Gamma$-K$_{\mathrm{W}}$, $\Gamma$-K$_{\mathrm{Mo}}$, and $\gamma$-$\kappa$. From this analysis, we calculate the twist angle to $\Theta=9.4\pm1.5^\circ$, which corresponds to a moiré periodicity of $\left|R_{\rm moire}\right|=$1.96$\pm$0.28~nm in real-space. The error margins are extracted from the error of the two-dimensional Gaussian fit and the subsequent error propagation for the calculation of the twist angle.

\vspace{1cm}

\begin{figure*}[hbt!]
    \centering
    \includegraphics[width=\linewidth]{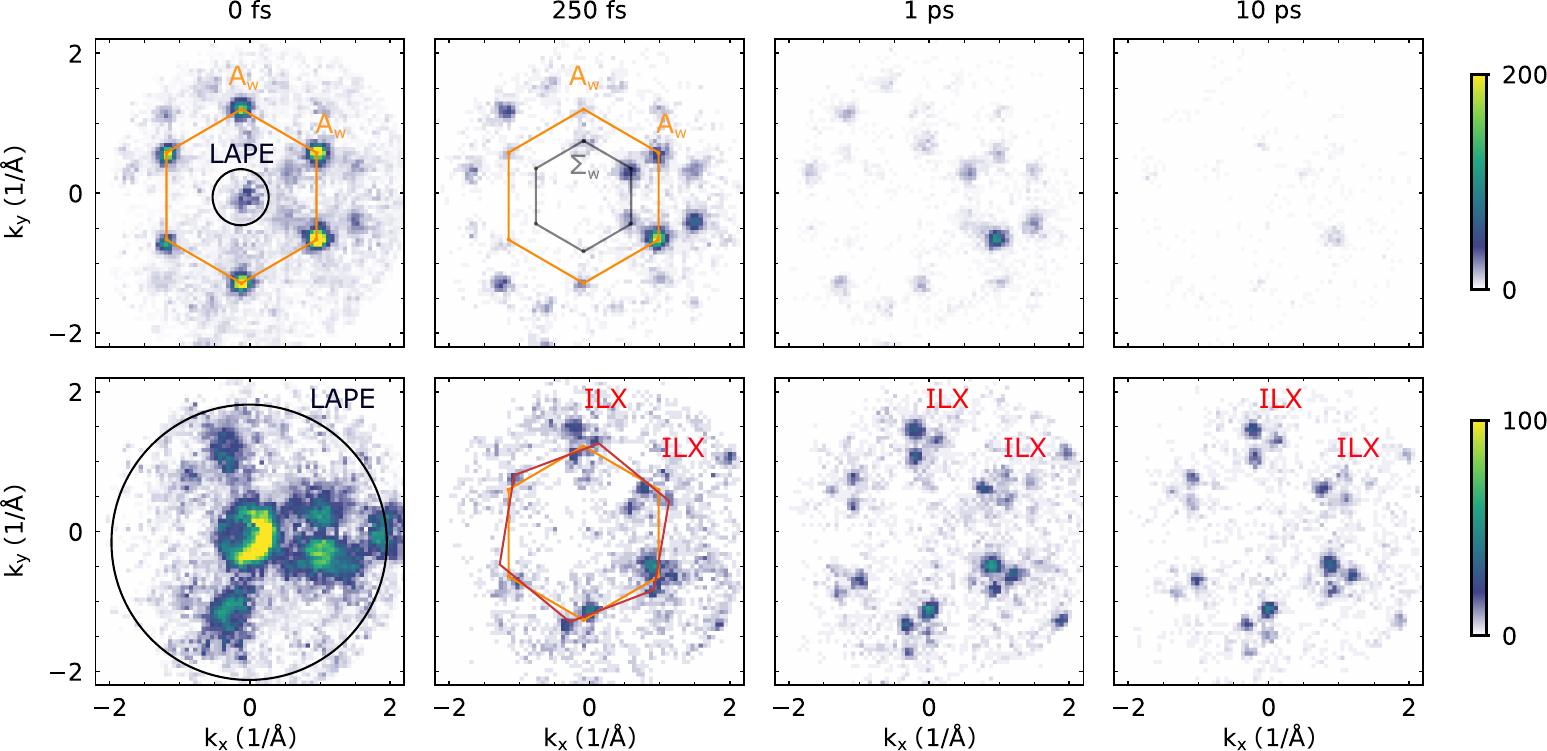}
    \caption{\textbf{Selected momentum maps of the intralayer excitons (top row) and the ILX (bottom row).}
    The momentum fingerprints of the excitons can be observed on the femto- to picosecond timescale (same data as shown in Fig. 3 of the main text). 
    At 0~fs, the bright $A_{\mathrm{W}}$-exciton, as well as signatures of the laser-assisted photoelectric effect are resolved. With increasing pump-probe delay, the dark $\Sigma_{\mathrm{W}}$-exciton and the ILX are formed. At 10~ps, only vanishing signal ob the intralayer excitons is observed, while the ILX signal persists. }
    \label{time_resolved_imaging}
\end{figure*}

\begin{figure*}[hbt!]
    \centering
    \includegraphics[width=.8\linewidth]{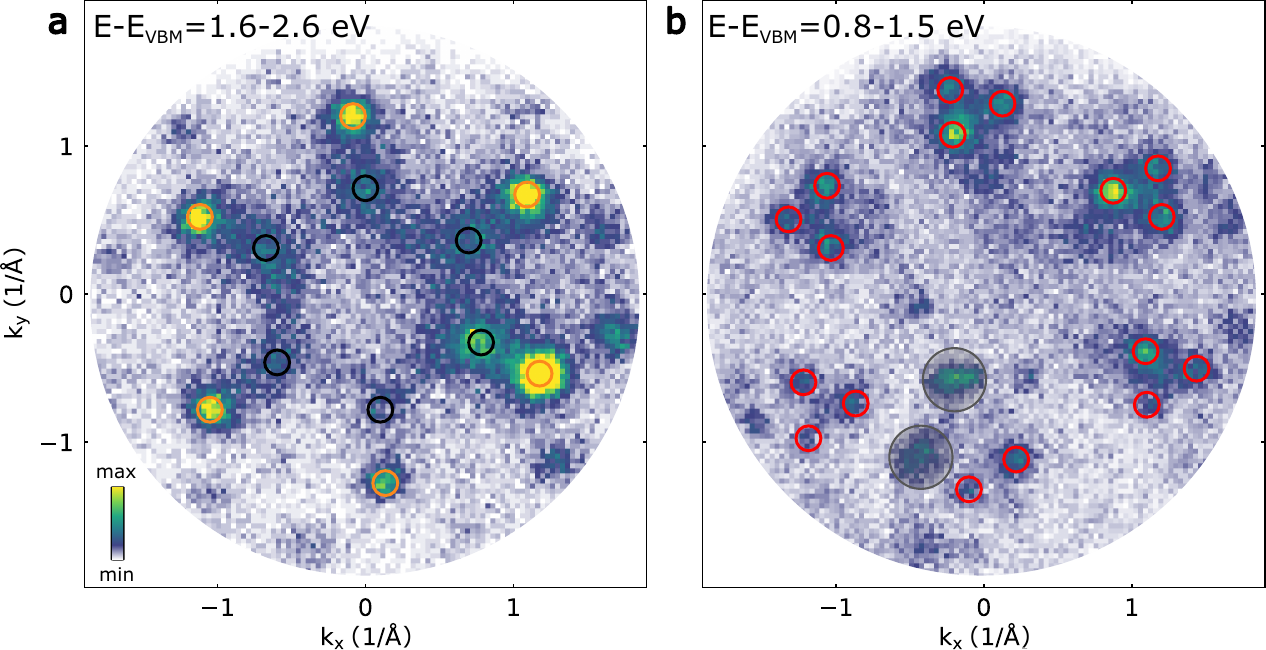}
    \caption{\textbf{Selected regions of interest for the analysis of the ILX formation dynamics shown in Fig.~2c}.
    \textbf{a} The A$_{\mathrm{W}}$- and the $\Sigma_{\mathrm{W}}$-exciton regions of interest are indicated by orange and black circles, respectively. The depicted momentum map is integrated over an energy window of 1.6 to 2.6~eV above the valence band maximum of WSe$_2$ and for all measured pump probe delays.
    \textbf{b} The ILX regions of interest are indicated by red circles. The momentum map is integrated over an energy window of 0.8 to 1.5~eV above the valence band maximum of WSe$_2$ and for all measured pump probe delays. The grey shaded areas indicate artefacts of our detector.
    } 
    \label{Rois}
\end{figure*}

\noindent\textbf{Extended time-resolved momentum microscopy data of the WSe$_2$/MoS$_2$ heterostructure.} In Fig.~2 of the main text, we analyse the formation dynamics of the ILX based on the multidimensional momentum microscopy data. A more detailed representation of the momentum maps in which we identify the intra- and interlayer excitons is shown in extended Fig.~\ref{time_resolved_imaging}. In addition, in extended Fig.~\ref{Rois}, we indicate the regions of interest that are used for the data evaluation in Fig.~2c.
\vspace{0.5cm}

\vspace{0.5cm}

\vspace{1cm}

\noindent\textbf{Construction of the moiré mini Brillouin zone (mBZ).} The moiré lattice vector $G^n_{\rm M}=C_3^n (G_{\mathrm{W}}^n - G_{\mathrm{Mo}}^n)$ builds up the moiré mBZ (red hexagon in Fig. 3d). G$^n\rm _{W}$ and G$^n\rm _{Mo}$ are the lattice vectors of the monolayers, and $C_3^n$ describes the three-fold rotational symmetry of the K valleys. The excitonic band structure of the ILX can then be described within the mBZ~\cite{Brem20nanolett, Wu18prb, Alexeev19nat}, whereas the mBZ center $\gamma$ ($\gamma$') overlaps with the K$_{\mathrm{W}}$ (K'$_{\mathrm{W}}$) valley (the hole location). The former $K_{\mathrm{Mo}}$ valleys (the electron locations) then coincide with the mBZ boundaries, i.e., on the $\kappa$ ($\kappa$') valleys.

\vspace{1cm}

\noindent\textbf{Correction of rigid band shifts.} In the time-resolved measurements, we observe an energetic shift of the momentum-integrated energy-distribution-curves on the order of up to 70~meV. The energy-shift is only observed in a small delay-window around time zero, i.e., when the pump and probe pulses are in temporal overlap. A true band gap renormalization due to a high density of excitons is reported to be observed on the picosecond timescale~\cite{Liu19prl} and can thus be excluded. In contrast, both space-charge and the surface photovoltage effects can show such signatures~\cite{Madeo20sci, Schonhense21rsi} but cannot be differentiated within our data. For the analysis shown in Fig. 2a and 2b of the main text, the energy-shift is corrected. Note that for the analysis of the ILX momentum-fingerprint, the energy-shift is irrelevant as the data is analyzed on the picosecond timescale. 
\vspace{0.5cm}

%\begin{figure*}[hbt!]
 %   \centering
  %  \includegraphics[width=\linewidth]{fig/supple/fig7/time_trace_only_spectrum_shift_v7.pdf}
   % \caption{\textbf{Correction of the rigid energy shift for various pump-probe delays.} In the raw %momentum microscopy data, we observe a rigid energetic shift of the full photoemission spectrum;
%    \textbf{a} the shift is centered around temporal overlap of the pump and the probe laser pulse and is at most $\mathrm{E}_{\mathrm{Peak}}-\mathrm{E}_{\mathrm{VBM}}$=70~meV, where $\mathrm{E}_{\mathrm{Peak}}$ is the energetic position of the photoemission signature labelled with red arrow in the inset of \textbf{b}. 
%    \textbf{b} The rigid band shift is calibrated on the peak position $\mathrm{E}_{\mathrm{Peak}}$.}
%    \label{correction_tr_data_set}
%\end{figure*}

\vspace{1cm}

\noindent\textbf{Real-space reconstruction.} Following the plane-wave model for photoemission, the measured ARPES intensity $I(\vec{k})$ can be expressed as
\begin{equation}
    I(\vec{k}) = |\vec{A}\cdot \vec{k}|^2~|FT\{\psi(\vec{r})\}|^2~\delta(E_b + E_{kin} + \Phi - \hbar \omega),
    \label{eq:plane_wave_model}
\end{equation}
which includes the Fourier transform of the real-space electronic wave function $\psi(\vec{r})$, a polarization factor $|\vec{A}\cdot \vec{k}|^2$ which depends on the vector potential $\vec{A}$ of the incident radiation and electron momentum $\vec{k}$, and finally a Dirac delta function which ensures conservation of energy. Because the electronic contribution to the excitonic quasiparticle is confined to a single MoS$_2$ monolayer, we can treat it similar to how is done for orbital tomography of molecular orbitals in planar aromatic molecules~\cite{Puschnig09sci, jansen_efficient_2020}: the wave function is assumed to be thin in the vertical dimension and photoemission is therefore assumed to be independent of the out-of-plane momentum $k_z$. We apply a symmetrization with respect to $\pm$120\textdegree{} rotation to match to the rotational symmetry of the moiré pattern, which eliminates the influence of the polarization factor and furthermore reduces our sensitivity to spurious local intensity variations which may arise due to sub-optimal alignment of the momentum microscope (insets in Fig. 4a and 4b of the main text). Finally, we subtract a weak background signal and set any resulting negative values to zero. 

To reconstruct accurate electronic wave functions, we furthermore need to separate contributions arising from the two spin-valley orientations in this system. Therefore, we apply a mask to select only the K or K' valleys and their direct vicinity. These masks are indicated by a gray overlay on the data in the insets of Fig.~4a and 4b. In order to reconstruct the real-space probability density from the resulting momentum maps, we only need to determine the phase pattern associated with the photoemission intensity.

For the intralayer A\textsubscript{W}-exciton, we impose a flat phase profile which is constant over the intensity distribution at each K (K') valley~\cite{Dong20naturalsciences}. For the more-complicated ILX, we base our phase retrieval approach on the following set of assumptions: First, we expect that each intensity spike in the momentum map has a single, flat phase associated with it. Furthermore, we assume that the spatial arrangement of the probability density is compact; in a pixel basis, this is equivalent to stating that the number of non-zero pixels is small. Using a sparsity-driven phase retrieval algorithm~\cite{jansen_efficient_2020}, we have verified that these requirements imply a flat, zero phase profile, which then yields the real-space reconstructions shown in Fig.~4a.

It is worth noting that based on the femtosecond momentum microscopy measurement, it is not possible to exclude a different true phase profile. Nevertheless, the zero phase assignment is instructive because it provides a lower bound on the spatial extent of the wave function. Similarly, we have not accounted for the limited resolution of the momentum microscope. As a lower momentum resolution would lead to a broadening of the bright spots in the momentum map and therefore a decrease in the spatial extent of the reconstruction, this again implies that the presented wave function reconstructions provide a lower bound on the spatial extent.

\vspace{1cm}

\noindent\textbf{Excluding final-state umklapp scattering.} Periodic superstructures can be imprinted onto momentum-resolved photoelectron spectra by the formation of replica signatures of the zero-order photoemission spectral weight. For the interpretation of these features, it is necessary to differentiate between (i) umklapp scattering occurring in the final states and (ii) a true modulation of the electronic band structure with the superlattice, i.e., the renormalization of the band structure. For the ILX, we can exclude the dominant contribution of final-state umklapp scattering: In a naive picture without the modulation of the ILX by the moiré potential, for the ILX, we would expect to detect photoelectrons originating from the K$_{\mathrm{Mo}}$ and K'$_{\mathrm{Mo}}$ valleys (i.e., from the valleys where the electron contribution to the ILX is localized, cf. Fig. 3c: corners of the brown Brillouin zone). In consequence, if we would only detect final-state umklapp processes from this ILX signature, we would expect to observe six photoemission spectral features centered around these K$_{\mathrm{Mo}}$ and K'$_{\mathrm{Mo}}$ valleys, but shifted by the moiré lattice vector $G_{M}^n$ (black arrows in Fig. 3d). Such scattering is observed, for example, in photoemission momentum distributions observed from graphene grown on Ir(111)~\cite{Pletikosic09prl}. Crucially, we do not observe six replica features. Rather, we observe only two prominent features which are shifted by the $G_{M}^n$ lattice vectors from the original K$_{\mathrm{Mo}}$ or K'$_{\mathrm{Mo}}$ valley (black arrows in Fig.~3d). As discussed in the main text, we instead explain the threefold-symmetric momentum fingerprint within the moiré mBZ of the ILX to be caused by the excitonic band structure of the ILX (red hexagon in Fig. 3d). We note that, although it is not possible to completely rule out a contribution of final-state umklapp processes, for the present study and its conclusions, umklapp scattering is not relevant.